\documentclass[english,preprint,nofootinbib]{revtex4}%
\usepackage[colorlinks=true,urlcolor=black,linkcolor=black,citecolor=black]%
{hyperref}
\usepackage[T1]{fontenc}
\usepackage[latin9]{inputenc}
\usepackage{amsmath}
\usepackage{amssymb}
\usepackage{amsfonts}
\usepackage{babel}
\usepackage{graphicx}%
\providecommand{\U}[1]{\protect\rule{.1in}{.1in}}
\makeatletter
\@ifundefined{textcolor}{}
{
\definecolor{BLACK}{gray}{0}
 \definecolor{WHITE}{gray}{1}
 \definecolor{RED}{rgb}{1,0,0}
 \definecolor{GREEN}{rgb}{0,1,0}
 \definecolor{BLUE}{rgb}{0,0,1}
 \definecolor{CYAN}{cmyk}{1,0,0,0}
 \definecolor{MAGENTA}{cmyk}{0,1,0,0}
 \definecolor{YELLOW}{cmyk}{0,0,1,0}
 }
\makeatother
\begin{document}
\title{Brief review on higher spin black holes}
\author{Alfredo Pérez$^{1}$, David Tempo$^{1}$, Ricardo Troncoso$^{1,2}$}
\email{aperez, tempo, troncoso@cecs.cl}
\affiliation{$^{1}$Centro de Estudios Científicos (CECs), Casilla 1469, Valdivia, Chile}
\affiliation{$^{2}$Universidad Andrés Bello, Av. República 440, Santiago, Chile.}
\preprint{CECS-PHY-14/01}

\begin{abstract}
We review some relevant results in the context of higher spin black holes in
three-dimensional spacetimes, focusing on their asymptotic behaviour and
thermodynamic properties. For simplicity, we mainly discuss the case of
gravity nonminimally coupled to spin-$3$ fields, being nonperturbatively
described by a Chern-Simons theory of two independent $sl\left(
3,\mathbb{R}\right)  $ gauge fields. Since the analysis is particularly
transparent in the Hamiltonian formalism, we provide a concise discussion of
their basic aspects in this context; and as a warming up exercise, we briefly
analyze the asymptotic behaviour of pure gravity, as well as the BTZ black
hole and its thermodynamics, exclusively in terms of gauge fields. The
discussion is then extended to the case of black holes endowed with higher
spin fields, briefly signaling the agreements and discrepancies found through
different approaches. We conclude explaining how the puzzles become resolved
once the fall off of the fields is precisely specified and extended to include
chemical potentials, in a way that it is compatible with the asymptotic
symmetries. Hence, the global charges become completely identified in an
unambiguous way, so that different sets of asymptotic conditions turn out to
contain inequivalent classes of black hole solutions being characterized by a
different set of global charges.

\end{abstract}
\maketitle

\section{Introduction}

Fundamental particles of spin greater than two are hitherto unknown, which
from a purely theoretical point of view, appears to agree with the widespread
belief that massless fields of spin $s>2$ are doomed to suffer from
inconsistencies. Indeed, the lore is reflected through a well-known claim in
the context of supergravity (see e.g., \cite{PTT-SUGRA}), which asserts that
the maximum number of local supersymmetries is bounded by eight; otherwise,
since the supersymmetry generators act as raising or lowering operators for
spin, a supermultiplet would contain fields of spin greater than two. In turn,
through the Kaluza-Klein mechanism, this also sets an upper bound on the
spacetime dimension to be at most eleven. The supposed inconsistency of higher
spin fields relies on solid no-go theorems (see \cite{PTT-BBS} for a good
review about this subject). In particular, it is worth mentioning the result
of Aragone and Deser \cite{PTT-Aragone-Deser}, which states that the higher
spin gauge symmetries of the free theory around flat spacetime, cannot be
preserved once the field is minimally coupled to gravity.

A consistent way to circumvent the incompatibility of higher spin gauge
symmetries with interactions was pioneered by Vasiliev \cite{PTT-VV1}%
,\ \cite{PTT-VV2}, who was able to formulate the field equations for a whole
tower of nonminimally coupled fields of spin $s=0$, $1$, $2$, ..., $\infty$,
in presence of a cosmological constant (For recent reviews see e.g.,
\cite{PTT-ReviewHS1-1}, \cite{PTT-ReviewHS1-2}). It is worth pointing out
that, since the hypotheses of the Coleman-Mandula theorem are not fulfilled by
Vasiliev theory, spacetime and gauge symmetries become inherently mixed in an
unaccustomed form \cite{PTT-Vasiliev-Paros}. It then goes without saying that
the very existence of Vasiliev theory, naturally suggests a possible
reformulation of supergravity theories from scratch, which would may in turn
elucidate new alternative approaches to strings and M-theory. Indeed, in
eleven dimensions and in presence of a negative cosmological constant, a
supergravity theory that shares some of these features, as the mixing of
spacetime and gauge symmetries, is known to exist \cite{PTT-TZ}.

In order to gain some insights about this counterintuitive subject, one may
instead follow the less ambitious approach of finding a simpler set up that
still captures some of the relevant features that characterize the dynamics of
higher spin fields. In this sense, the three-dimensional case turns out to be
particularly appealing, since the dynamics is described through a standard
field theory with a Chern-Simons action \cite{PTT-3D-2},\ \cite{PTT-3D-3},
\cite{PTT-3D-4}. The generic theory can be further simplified, since it admits
a consistent truncation to the case of a finite number of nonpropagating
fields with spin $s=2$, $3$, ..., $N$. Hence the simplest case with the
desired properties corresponds to $N=3$, so that the theory describes gravity
with negative cosmological constant, nonminimally coupled to an interacting
spin-three field. The remarkable simplification of the theory then allows the
possibility of finding different classes of exact black hole solutions endowed
with a nontrivial spin-three field, as the ones in \cite{PTT-GK},
\cite{PTT-CM}, and \cite{PTT-HPTT}, respectively. However, despite the
simplicity of these solutions, the subject has not been free of controversy,
mainly due to the puzzling discrepancies that have been found in the
characterization of their global charges and their entropy.

\bigskip

The purpose of this brief review, is overviewing some of the relevant results
about this ongoing subject, as well as explaining how the apparent tension
between different approaches is fully resolved once the chemical potentials
are suitably identified along the lines of \cite{PTT-HPTT}, \cite{PTT-BHPTT},
so that the asymptotic symmetries, and hence the global charges, are
completely characterized in an unambiguous way.

It is worth highlighting that the action principle in terms of the metric and
the spin-3 field is currently known as a weak field expansion of the spin-3
field up to quadratic order \cite{PTT-CFPT2}. Thus, in order to deal with the
full nonperturbative treatment of the higher spin black hole solutions, it
turns out to be useful to describe them only in terms of gauge fields and the
topology of the manifold, without making any reference neither to the metric
nor to the spin-3 field.

Since the analysis becomes particularly transparent in the Hamiltonian
formalism, in the next section we concisely discuss some of their basic
aspects in the context of Chern-Simons theories in three dimensions. As a
useful warming up exercise, in section \ref{PTT-Pure grav}, the asymptotic
behaviour of pure gravity with negative cosmological constant
\cite{PTT-Brown-Henneaux}, as well as the BTZ black hole \cite{PTT-BTZ},
\cite{PTT-BHTZ} and its thermodynamics, are briefly analyzed exclusively in
terms of gauge fields. Section \ref{PTT-HSG3D} is devoted to the case of
gravity coupled to spin-3 fields, including the asymptotic behaviour described
in \cite{PTT-Henneaux-Rey}, \cite{PTT-CFPT1}, the higher spin black hole
solution of \cite{PTT-GK}, \cite{PTT-AGKP}, and its thermodynamics
\cite{PTT-PTT1}, \cite{PTT-PTT2}, briefly signaling the agreements and
discrepancies found through different approaches. We conclude with section
\ref{PTT-Puzzles}, where it is explained how these puzzling differences become
fully resolved once the fall off of the fields is precisely specified, so that
different sets of asymptotic conditions turn out to contain inequivalent
classes of black hole solutions \cite{PTT-HPTT}, \cite{PTT-BHPTT} being
characterized by a different set of global charges.

\section{Basic aspects and Hamiltonian formulation of Chern-Simons theories in
three dimensions}

In three-dimensional spacetimes, gauge theories described by a Chern-Simons
action are much simpler than their corresponding Yang-Mills analogues, in the
sense that less structure is required in order to formulate them. Indeed, the
manifold $M$, locally described by a set of coordinates $x^{\mu}$, is only
endowed with a gauge field $A=A_{\mu}^{I}T_{I}dx^{\mu}$, where $T_{I}$ stand
for the generators of a Lie algebra $\mathfrak{g}$, which is assumed to admit
an invariant nondegenerate bilinear form $g_{IJ}=\left\langle T_{I}%
,T_{J}\right\rangle $. These ingredients are enough to construct the action,
given by%
\begin{equation}
I_{CS}\left[  A\right]  =\frac{k}{4\pi}\int_{M}\left\langle AdA+\frac{2}%
{3}A^{3}\right\rangle \ , \label{PTT-ICS}%
\end{equation}
where $k$ is a constant, and wedge product between forms has been assumed.
Consequently, the action does not require the existence of a spacetime metric,
but it is sensitive to the topology of $M$. The field equations imply the
vanishing of curvature, i.e., $F=dA+A^{2}=0$, so that the connection becomes
locally flat on shell, and then the theory is devoid of local propagating
degrees of freedom. Note that the action (\ref{PTT-ICS}) is already in
Hamiltonian form. Indeed, if the topology of $M$ is of the form $M=\Sigma
\times%
\mathbb{R}
$, where $\Sigma$ stands for the spacelike section, the connection splits as
$A=A_{i}dx^{i}+A_{t}dt$, and hence the action (\ref{PTT-ICS}) reduces to%
\begin{equation}
I_{H}=-\frac{k}{4\pi}\int_{M}dtd^{2}x\varepsilon^{ij}\left\langle A_{i}\dot
{A}_{j}-A_{t}F_{ij}\right\rangle \ , \label{PTT-ICSH}%
\end{equation}
up to a boundary term. It is then apparent that $A_{i}$ correspond to the
dynamical fields, whose Poisson brackets are given by $\left\{  A_{i}%
^{I}\left(  x\right)  ,A_{j}^{J}\left(  x^{\prime}\right)  \right\}
=\frac{2\pi}{k}g^{IJ}\varepsilon_{ij}\delta\left(  x-x^{\prime}\right)  $,
while $A_{t}$ become Lagrange multipliers associated to the constraints
$G=\frac{k}{4\pi}\varepsilon^{ij}F_{ij}$. Then, the smeared generator of the
gauge transformations reads%
\[
G\left(  \Lambda\right)  =\int_{\Sigma}d^{2}x\left\langle \Lambda
G\right\rangle \ ,
\]
so that $\delta A_{i}=\left\{  A_{i},G\left(  \Lambda\right)  \right\}
=\partial_{i}\Lambda+\left[  A_{i},\Lambda\right]  $ (see, e.g.,
\cite{PTT-Balachandran}, \cite{PTT-Banados-Q}, \cite{PTT-Carlip-Q}). However,
when $\Sigma$ has a boundary, according to the Regge-Teitelboim approach
\cite{PTT-Regge-Teitelboim}, the generator of the gauge transformations has to
be improved by a boundary term $Q\left(  \Lambda\right)  $, i.e.,
\begin{equation}
\tilde{G}\left(  \Lambda\right)  =G\left(  \Lambda\right)  +Q\left(
\Lambda\right)  \ , \label{PTT-improved-G}%
\end{equation}
being such that its functional variation is well-defined everywhere. This
implies that the variation of the the conserved charge associated to an
asymptotic gauge symmetry, generated by a Lie algebra valued parameter
$\Lambda$, is determined by the dynamical fields at a fixed time slice at the
boundary, which reads%
\begin{equation}
\delta Q\left(  \Lambda\right)  =-\frac{k}{2\pi}\int_{\partial\Sigma
}\left\langle \Lambda\delta A_{\theta}\right\rangle d\theta\ ,
\label{PTT-deltaqeta}%
\end{equation}
where $\partial\Sigma$ stands for the boundary of the spacelike section
$\Sigma$.

The transformation law of the Lagrange multipliers, $\delta A_{t}=\partial
_{t}\Lambda+\left[  A_{t},\Lambda\right]  $, is then recovered requiring the
improved action to be invariant under gauge transformations. Note that
on-shell, by virtue of the identity $\mathcal{L}_{\xi}A_{\mu}=\nabla_{\mu
}\left(  \xi^{\nu}A_{\nu}\right)  +\xi^{\nu}F_{\nu\mu}$, diffeomorphisms
$\delta_{\xi}A_{\mu}=-\mathcal{L}_{\xi}A_{\mu}$ are equivalent to gauge
transformations with parameter $\Lambda=-\xi^{\mu}A_{\mu}$, and hence, the
variation of the generator of an asymptotic symmetry spanned by an asymptotic
killing vector $\xi^{\mu}$, reads%
\begin{equation}
\delta Q\left(  \xi\right)  =\frac{k}{2\pi}\int_{\partial\Sigma}\xi^{\mu
}\left\langle A_{\mu}\delta A_{\theta}\right\rangle d\theta\ .
\label{PTT-deltaqchi}%
\end{equation}
This means that the variation of the total energy of the system, which takes
into account the contribution of all the constraints, is given by%
\begin{equation}
\delta E=\delta Q\left(  \partial_{t}\right)  =\frac{k}{2\pi}\int
_{\partial\Sigma}\left\langle A_{t}\delta A_{\theta}\right\rangle d\theta\ .
\label{PTT-deltaE}%
\end{equation}

It should be stressed that the whole canonical structure only makes sense
provided the variation of the canonical generators can be integrated. This can
be generically done once a precise set of asymptotic conditions is specified,
which in turn determines the asymptotic symmetries. This will be explicitly
discussed in the next section for the case of pure gravity with negative
cosmological constant, as well as in section \ref{PTT-HSG3D}, and further
elaborated in section \ref{PTT-Puzzles} in the case of gravity coupled to a
spin-3 field.

\section{General Relativity with negative cosmological constant in three
dimensions}

\label{PTT-Pure grav}

As it was shown in \cite{PTT-AT}, \cite{PTT-W} General Relativity in vacuum
can be described in terms of a Chern-Simons action. In the case of negative
cosmological constant the corresponding Lie algebra is of the form
$\mathfrak{g}=\mathfrak{g}_{+}+\mathfrak{g}_{-}$, where $\mathfrak{g}_{\pm}$
stand for two independent copies of $sl\left(  2,\mathbb{R}\right)  $, which
will be assumed to be described by the same set of matrices $L_{i}$, with
$i=-1,0,1$, given by
\begin{equation}
L_{-1}=%
\begin{pmatrix}
0 & 0\\
1 & 0
\end{pmatrix}
\quad;\quad L_{0}=%
\begin{pmatrix}
-\frac{1}{2} & 0\\
0 & \frac{1}{2}%
\end{pmatrix}
\quad;\quad L_{1}=%
\begin{pmatrix}
0 & -1\\
0 & 0
\end{pmatrix}
\;, \label{PTT-sl(2,R)-MR}%
\end{equation}
so that the $sl\left(  2,\mathbb{R}\right)  $ algebra reads%
\begin{equation}
\left[  L_{i},L_{j}\right]  =\left(  i-j\right)  L_{i+j}\;.
\end{equation}
The connection then splits in two independent $sl\left(  2,\mathbb{R}\right)
$-valued gauge fields, according to $A=A^{+}+A^{-}$, while the invariant
nondegenerate bilinear form is chosen such that the action (\ref{PTT-ICS})
reduces to%
\begin{equation}
I=I_{CS}\left[  A^{+}\right]  -I_{CS}\left[  A^{-}\right]  \ ,
\end{equation}
so that the bracket now corresponds to just the trace, i.e., in the
representation of (\ref{PTT-sl(2,R)-MR}), $\left\langle \cdots\right\rangle
=\mathrm{tr}\left(  \cdots\right)  $, and the level is fixed by the AdS radius
and the Newton constant as $k=\frac{l}{4G}$. The link between the gauge fields
and spacetime geometry is made through%
\begin{equation}
A^{\pm}=\omega\pm\frac{e}{l}\ , \label{PTT-Amn-GR}%
\end{equation}
where $\omega$ and $e$ correspond to the spin connection and the dreibein,
respectively. The field equations, $F^{\pm}=0$, then imply that the spacetime
curvature is constant and the torsion vanishes, while the metric is recovered
from%
\begin{equation}
g_{\mu\nu}=2\mathrm{tr}\left(  e_{\mu}e_{\nu}\right)  \ , \label{PTT-gmunu}%
\end{equation}
which is manifestly invariant under the diagonal subgroup of $SL\left(
2,\mathbb{R}\right)  \times SL\left(  2,\mathbb{R}\right)  $, that corresponds
to the local Lorentz transformations. Note that diffeomorphisms can always be
expressed in terms of the remaining gauge symmetries.

\subsection{Brown-Henneaux boundary conditions}

\label{PTT-Brown-Henneaux-Section}

As explained in \cite{PTT-CHvD}, the asymptotic behaviour of gravity with
negative cosmological constant, as originally described by Brown and Henneaux
\cite{PTT-Brown-Henneaux}, can be readily formulated in terms of the gauge
fields $A^{\pm}$. The gauge can be chosen such that the radial dependence is
entirely captured by the group elements
\begin{equation}
g_{\pm}=e^{\pm\rho L_{0}}\ , \label{PTT-gmn}%
\end{equation}
so that the asymptotic form of the connections is given by%
\begin{equation}
A^{\pm}=g_{\pm}^{-1}a^{\pm}g_{\pm}+g_{\pm}^{-1}dg_{\pm}\ ,
\label{PTT-Acontuti}%
\end{equation}
where $a^{\pm}=a_{\theta}^{\pm}d\theta+a_{t}^{\pm}dt$, read%
\begin{equation}
a^{\pm}=\pm\left(  L_{\pm1}-\frac{2\pi}{k}\mathcal{L}_{\pm}L_{\mp1}\right)
dx^{\pm}\ , \label{PTT-BH-BCs}%
\end{equation}
with $x^{\pm}=\frac{t}{l}\pm\theta$, and the functions $\mathcal{L}_{\pm}$
depend only on time and the angular coordinate.

The asymptotic form of the dynamical fields $a_{\theta}^{\pm}$ is preserved
under gauge transformations, $\delta a_{\theta}^{\pm}=\partial_{\theta}%
\Lambda^{\pm}+\left[  a_{\theta}^{\pm},\Lambda^{\pm}\right]  $, generated by%
\begin{equation}
\Lambda^{\pm}\left(  \varepsilon_{\pm}\right)  =\varepsilon_{\pm}L_{\pm1}%
\mp\varepsilon_{\pm}^{\prime}L_{0}+\frac{1}{2}\left(  \varepsilon_{\pm
}^{\prime\prime}-\frac{4\pi}{k}\varepsilon_{\pm}\mathcal{L}_{\pm}\right)
L_{\mp1}\ , \label{PTT-Lambda-GR}%
\end{equation}
where $\varepsilon_{\pm}$ are arbitrary functions of $t$, $\theta$, provided
the functions $\mathcal{L}_{\pm}$ transform as%
\begin{equation}
\delta\mathcal{L}_{\pm}=\varepsilon_{\pm}\mathcal{L}_{\pm}^{\prime
}+2\mathcal{L}_{\pm}\varepsilon_{\pm}^{\prime}-\frac{k}{4\pi}\varepsilon_{\pm
}^{\prime\prime\prime}\ . \label{PTT-deltaL-GR}%
\end{equation}
Hereafter, prime denotes the derivative with respect to $\theta$. Furthermore,
requiring the components of the gauge fields along time, $a_{t}^{\pm}$, to be
mapped into themselves under the same gauge transformations, together with the
transformation laws in (\ref{PTT-deltaL-GR}), implies that the functions
$\mathcal{L}_{\pm}$ and the parameters $\varepsilon_{\pm}$ are chiral, i.e.,%
\begin{equation}
\partial_{\mp}\mathcal{L}_{\pm}=0\ \ ,\ \ \partial_{\mp}\varepsilon_{\pm}=0\ .
\label{PTT-FE+CC-GR}%
\end{equation}
Note that the first condition in (\ref{PTT-FE+CC-GR}) means that the field
equations have to be fulfilled in the asymptotic region.

Consequently, according to (\ref{PTT-deltaqeta}), the variation of the
canonical generators associated to the asymptotic gauge symmetries generated
by $\Lambda=\Lambda^{+}+\Lambda^{-}$, in this case reduces to%
\begin{equation}
\delta Q\left(  \Lambda\right)  =\delta Q_{+}\left(  \Lambda^{+}\right)
-\delta Q_{-}\left(  \Lambda^{-}\right)  \ ,
\end{equation}
with%
\begin{equation}
\delta Q_{\pm}\left(  \Lambda^{\pm}\right)  =-\frac{k}{2\pi}\int\left\langle
\Lambda^{\pm}\delta a_{\theta}^{\pm}\right\rangle d\theta=-\int\varepsilon
_{\pm}\delta\mathcal{L}_{\pm}d\theta\ ,
\end{equation}
which can be readily integrated as%
\begin{equation}
Q_{\pm}\left(  \Lambda^{\pm}\right)  =-\int\varepsilon_{\pm}\mathcal{L}_{\pm
}d\theta\ . \label{PTT-Q-Brown-Henneaux}%
\end{equation}
Therefore, since the canonical generators fulfill $\delta_{\Lambda_{1}%
}Q\left[  \Lambda_{2}\right]  =\left\{  Q\left[  \Lambda_{2}\right]  ,Q\left[
\Lambda_{1}\right]  \right\}  $, their algebra can be directly obtained by
virtue of (\ref{PTT-deltaL-GR}), which reduces to two copies of the Virasoro
algebra with the same central extension $c=\frac{3l}{2G}$
\cite{PTT-Brown-Henneaux}. Expanding in Fourier modes, according to
$\mathcal{L}=\frac{1}{2\pi}\sum_{m}\mathcal{L}_{m}e^{im\theta}$, the algebra
explicitly reads%
\begin{equation}
i\left\{  \mathcal{L}_{m},\mathcal{L}_{n}\right\}  =\left(  m-n\right)
\mathcal{L}_{m+n}+\frac{k}{2}m^{3}\delta_{m+n,0}\ ,
\end{equation}
for both copies.

\subsection{BTZ black hole and its thermodynamics}

The asymptotic conditions described above, manifestly contain the BTZ black
hole solution \cite{PTT-BTZ}, \cite{PTT-BHTZ}, being described by%
\begin{equation}
a^{\pm}=\pm\left(  L_{\pm1}-\frac{2\pi}{k}\mathcal{L}_{\pm}L_{\mp1}\right)
dx^{\pm}\ , \label{PTT-amn-BTZ}%
\end{equation}
when $\mathcal{L}_{\pm}$ are nonnegative constants. Indeed, by virtue of eqs.
(\ref{PTT-Amn-GR}) and (\ref{PTT-gmunu}), the spacetime metric is recovered in
normal coordinates:%
\begin{align}
ds^{2}  &  =l^{2}\left[  d\rho^{2}+\frac{2\pi}{k}\left(  \mathcal{L}%
_{+}\left(  dx^{+}\right)  ^{2}+\mathcal{L}_{-}\left(  dx^{-}\right)
^{2}\right)  \right. \nonumber\\
&  \left.  -\left(  e^{2\rho}+\frac{4\pi^{2}}{k^{2}}\mathcal{L}_{+}%
\mathcal{L}_{-}e^{-2\rho}\right)  dx^{+}dx^{-}\right]  \ .
\label{PTT-BTZ-metricNC}%
\end{align}
As shown in \cite{PTT-Carlip-Teiltelboim} (see also \cite{PTT-MS}), the
topology of the Euclidean black hole corresponds to $%
\mathbb{R}
^{2}\times S^{1}$, where $%
\mathbb{R}
^{2}$ stands for the one of the $\rho-\tau$ plane, and $\tau=-it$ is the
Euclidean time, fulfilling $0\leq\tau<\beta$, where $\beta=T^{-1}$ is the
inverse of the Hawking temperature. Since $%
\mathbb{R}
^{2}$ can be mapped into a disk through a conformal compactification, the
black hole topology is then equivalent to the one of a solid torus.

As explained in the introduction, and for later purposes, afterwards we will
perform the remaining analysis exclusively in terms of the gauge fields
(\ref{PTT-amn-BTZ}) and the topology of the manifold, without making any
reference to the spacetime metric.

The simplest gauge covariant object that is sensitive to the global properties
of the manifold turns out to be the holonomy of the gauge field around a
closed cycle $\mathcal{C}$, defined as%
\begin{equation}
\mathcal{H}_{\mathcal{C}}=P\exp\left(  \int_{\mathcal{C}}A_{\mu}dx^{\mu
}\right)  \ ,
\end{equation}
which is an element of the gauge group. Hence, since in this case the gauge
group corresponds to $SL\left(  2,%
\mathbb{R}
\right)  \times SL\left(  2,%
\mathbb{R}
\right)  $, the holonomy around $\mathcal{C}$\ is of the form $\mathcal{H}%
_{\mathcal{C}}=\mathcal{H}_{\mathcal{C}}^{+}\otimes\mathcal{H}_{\mathcal{C}%
}^{-}$, with%
\begin{equation}
\mathcal{H}_{\mathcal{C}}^{\pm}=P\exp\left(  \int_{\mathcal{C}}A_{\mu}^{\pm
}dx^{\mu}\right)  \ .
\end{equation}
As the topology of the manifold is the one of a solid torus, there are two
inequivalent classes of cycles: (I) the ones that wind around the handle, and
(II) those that do not. This means that the former ones are noncontractible,
while the latter can be continuously shrunk to a point. Then, the holonomies
along contractible cycles are trivial, i.e.,%
\begin{equation}
\mathcal{H}_{\mathcal{C}_{II}}^{\pm}=-\mathrm{1}\ , \label{PTT-Hii-BTZ}%
\end{equation}
where the negative sign is due to the fact that, according to
(\ref{PTT-sl(2,R)-MR}), we are dealing with the fundamental (spinorial)
representation of $SL\left(  2,%
\mathbb{R}
\right)  $; while the holonomies along noncontractible cycles $\mathcal{H}%
_{\mathcal{C}_{I}}^{\pm}$ are necessarily nontrivial. Indeed, it is easy to
verify that this is the case for the gauge fields that describe the BTZ black
hole (\ref{PTT-amn-BTZ}). For simplicity, we explicitly carry out the
computation in the static case, i.e., for $\mathcal{L}:=\mathcal{L}_{\pm}$,
since the inclusion of rotation is straightforward.

A simple noncontractible cycle in this case is parametrized by $\rho=\rho_{0}%
$, and $\tau=\tau_{0}$, with $\rho_{0}$, $\tau_{0}$ constants, so that the
corresponding holonomies around it read%
\begin{equation}
\mathcal{H}_{\theta}^{\pm}=e^{2\pi a_{\theta}^{\pm}}\ . \label{PTT-Hphi-BTZ}%
\end{equation}
These holonomies are then fully characterized, up to conjugacy by elements of
$SL\left(  2,%
\mathbb{R}
\right)  $, by the eigenvalues of $2\pi a_{\theta}^{\pm}$, given by%
\begin{equation}
\lambda_{\pm}^{2}=2\pi^{2}\mathrm{tr}\left[  \left(  a_{\theta}^{\pm}\right)
^{2}\right]  =\frac{8\pi^{3}}{k}\mathcal{L}\ ,
\end{equation}
and hence, since $\mathcal{L}$ is nonnegative, they are manifestly nontrivial.

Analogously, a simple contractible cycle is parametrized by $\rho=\rho_{0}$,
and $\theta=\theta_{0}$. Since the holonomies around this cycle are trivial,
the conditions in (\ref{PTT-Hii-BTZ}) reduce to%
\begin{equation}
\mathcal{H}_{\tau}^{\pm}=e^{\beta a_{\tau}^{\pm}}=e^{i\beta a_{t}^{\pm}%
}=-\mathrm{1}\ , \label{PTT-Htau-BTZ}%
\end{equation}
and since the cycle winds once, the eigenvalues of $i\beta a_{t}$ are given by
$\pm i\pi$, which equivalently implies that%
\begin{equation}
\beta^{2}\mathrm{tr}\left[  \left(  a_{t}^{\pm}\right)  ^{2}\right]  =2\pi
^{2}\ . \label{PTT-BetaEq-BTZ}%
\end{equation}
Therefore, the triviality of the holonomies around this cycle amounts to fix
the Euclidean time period as%
\begin{equation}
\beta=l\sqrt{\frac{\pi k}{2\mathcal{L}}}\ , \label{PTT-beta-BTZ}%
\end{equation}
in full agreement with the Hawking temperature.

Note that the variation of the total energy (\ref{PTT-deltaE}) in this case
reads%
\begin{equation}
\delta E=\frac{k}{2\pi}\int_{\partial\Sigma}\left(  \left\langle a_{t}%
^{+}\delta a_{\theta}^{+}\right\rangle -\left\langle a_{t}^{-}\delta
a_{\theta}^{-}\right\rangle \right)  d\theta=\frac{4\pi}{l}\delta
\mathcal{L\ },
\end{equation}
from which, by virtue of (\ref{PTT-beta-BTZ}) and the first law, implies that%
\begin{equation}
\delta S=\beta\delta E=\delta\left(  4\pi\sqrt{2\pi k\mathcal{L}}\right)  \ ,
\end{equation}
which means that the entropy can be expressed in terms of the global charges
(\ref{PTT-Q-Brown-Henneaux}), as%
\begin{equation}
S=4\pi\sqrt{2\pi k\mathcal{L}}\ . \label{PTT-S-BTZ}%
\end{equation}

\bigskip

The black hole entropy found in this way agrees with the standard result
obtained in the metric formalism. Indeed, according to (\ref{PTT-BTZ-metricNC}%
), in the static case the event horizon is located at $e^{2\rho_{+}}%
=\frac{2\pi}{k}\mathcal{L}$, so that its area is given by $A=4\pi l\sqrt
{\frac{2\pi}{k}\mathcal{L}}$, and hence (\ref{PTT-S-BTZ}) is equivalent to the
Bekenstein-Hawking formula $S=\frac{A}{4G}$.

\section{Higher spin gravity in 3D}

\label{PTT-HSG3D}

As explained in the introduction, gravity with negative cosmological constant,
nonminimally coupled to an interacting spin-$3$ field can be described in
terms of a Chern-Simons theory \cite{PTT-3D-2},\ \cite{PTT-3D-3},
\cite{PTT-3D-4}. The action is then of the form (\ref{PTT-ICS}), and as in the
case of pure gravity, the corresponding Lie algebra is of the form
$\mathfrak{g}=\mathfrak{g}_{+}+\mathfrak{g}_{-}$, but where now $\mathfrak{g}%
_{\pm}$ are enlarged to two independent copies of $sl\left(  3,\mathbb{R}%
\right)  $. Both copies of the algebra will be assumed to be spanned by the
same set of matrices $L_{i}$, $W_{m}$, with $i=-1,0,1$, and $m=-2,-1,0,1,2$,
given by (see e.g., \cite{PTT-CFPT1})%
\[
L_{-1}=%
\begin{pmatrix}
0 & -2 & 0\\
0 & 0 & -2\\
0 & 0 & 0
\end{pmatrix}
\quad;\quad L_{0}=%
\begin{pmatrix}
1 & 0 & 0\\
0 & 0 & 0\\
0 & 0 & -1
\end{pmatrix}
\quad;\quad L_{1}=%
\begin{pmatrix}
0 & 0 & 0\\
1 & 0 & 0\\
0 & 1 & 0
\end{pmatrix}
\;,
\]%
\begin{equation}
W_{-2}=%
\begin{pmatrix}
0 & 0 & 8\\
0 & 0 & 0\\
0 & 0 & 0
\end{pmatrix}
\quad;\quad W_{-1}=%
\begin{pmatrix}
0 & -2 & 0\\
0 & 0 & 2\\
0 & 0 & 0
\end{pmatrix}
\quad;\quad W_{0}=\frac{2}{3}%
\begin{pmatrix}
1 & 0 & 0\\
0 & -2 & 0\\
0 & 0 & 1
\end{pmatrix}
\;, \label{PTT-MR}%
\end{equation}%
\[
W_{1}=%
\begin{pmatrix}
0 & 0 & 0\\
1 & 0 & 0\\
0 & -1 & 0
\end{pmatrix}
\quad;\quad W_{2}=%
\begin{pmatrix}
0 & 0 & 0\\
0 & 0 & 0\\
2 & 0 & 0
\end{pmatrix}
\;,
\]
whose commutation relations read%
\begin{align}
\left[  L_{i},L_{j}\right]   &  =\left(  i-j\right)  L_{i+j}\;,\nonumber\\
\left[  L_{i},W_{m}\right]   &  =\left(  2i-m\right)  W_{i+m}%
\;,\label{PTT-sl(3,R)-algebra}\\
\left[  W_{m},W_{n}\right]   &  =-\frac{1}{3}\left(  m-n\right)  \left(
2m^{2}+2n^{2}-mn-8\right)  L_{m+n}\;,\nonumber
\end{align}
so that the subset of generators $L_{i}$ span the algebra $sl\left(  2,%
\mathbb{R}
\right)  $ in the so-called principal embedding.

The invariant nondegenerate bilinear form can also be chosen so that the
action (\ref{PTT-ICS}) reads%
\begin{equation}
I=I_{CS}\left[  A^{+}\right]  -I_{CS}\left[  A^{-}\right]  \ ,
\end{equation}
where $A^{\pm}$ stand for the gauge fields that correspond to both copies of
$sl\left(  3,\mathbb{R}\right)  $, and now the bracket is given by a quarter
of the trace in the representation of (\ref{PTT-MR}), i.e., $\left\langle
\cdots\right\rangle =\frac{1}{4}\mathrm{tr}\left(  \cdots\right)  $. As in the
case of pure gravity, the level is also chosen as $k=\frac{l}{4G}$.

It is useful to introduce a generalization of the dreibein and the spin
connection, which relate with the gauge fields according to%
\begin{equation}
A^{\pm}=\omega\pm\frac{e}{l}\ ,
\end{equation}
so that the spacetime metric and the spin-$3$ field can be recovered as%
\begin{equation}
g_{\mu\nu}=\frac{1}{2}\mathrm{tr}\left(  e_{\mu}e_{\nu}\right)  \ ;\ \varphi
_{\mu\nu\rho}=\frac{1}{3!}\mathrm{tr}\left(  e_{(\mu}e_{\nu}e_{\rho)}\right)
\ , \label{PTT-metric+spin3}%
\end{equation}
being manifestly invariant under the diagonal subgroup of $SL\left(
3,\mathbb{R}\right)  \times SL\left(  3,\mathbb{R}\right)  $, which
corresponds to an extension of the local Lorentz group. The remaining gauge
symmetries are then not only related to diffeomorphisms, but also with the
higher spin gauge transformations. It is worth pointing out that, since the
metric transforms in a nontrivial way under the action of the higher spin
gauge symmetries, some standard geometric and physical notions turn out to be
ambiguous, since they are no longer invariant. This last observation can be
regarded as an additional motivation to explore the physical properties of the
theory directly in terms of its original variables, given by the gauge fields
$A^{\pm}$.

\subsection{Asymptotic conditions with $W_{3}$ symmetries}

A consistent set of asymptotic conditions for the theory described above was
found in \cite{PTT-Henneaux-Rey}, \cite{PTT-CFPT1}. Using the gauge choice as
in \cite{PTT-CHvD}, the radial dependence can be completely absorbed by
$SL\left(  3,%
\mathbb{R}
\right)  $ group elements of the form (\ref{PTT-gmn}), so that the asymptotic
behaviour of the gauge fields can be written as in eq. (\ref{PTT-Acontuti}),
where $a^{\pm}$ are now given by%
\begin{equation}
a^{\pm}=\pm\left(  L_{\pm1}-\frac{2\pi}{k}\mathcal{L}_{\pm}L_{\mp1}-\frac{\pi
}{2k}\mathcal{W}_{\pm}W_{\mp2}\right)  dx^{\pm}\ , \label{PTT-amn-Standard W3}%
\end{equation}
and $\mathcal{L}_{\pm}$, $\mathcal{W}_{\pm}$ stand for arbitrary functions of
$t$, $\theta$. The asymptotic symmetries can then be readily found following
the same steps as in the case of pure gravity, previously discussed in section
\ref{PTT-Brown-Henneaux-Section}.

The asymptotic form of the fields $a_{\theta}^{\pm}$ is maintained under gauge
transformations generated by%
\begin{align}
\Lambda^{\pm}\left(  \varepsilon_{\pm},\chi_{\pm}\right)   &  =\varepsilon
_{\pm}L_{\pm1}+\chi_{\pm}W_{\pm2}\mp\varepsilon_{\pm}^{\prime}L_{0}\mp
\chi_{\pm}^{\prime}W_{\pm1}+\frac{1}{2}\left(  \varepsilon_{\pm}^{\prime
\prime}-\frac{4\pi}{k}\varepsilon_{\pm}\mathcal{L}_{\pm}+\frac{8\pi}%
{k}\mathcal{W}_{\pm}\chi_{\pm}\right)  L_{\mp1}\nonumber\\
&  -\left(  \frac{\pi}{2k}\mathcal{W}_{\pm}\varepsilon_{\pm}+\frac{7\pi}%
{6k}\mathcal{L}_{\pm}^{\prime}\chi_{\pm}^{\prime}+\frac{\pi}{3k}\chi_{\pm
}\mathcal{L}_{\pm}^{\prime\prime}+\frac{4\pi}{3k}\mathcal{L}_{\pm}\chi_{\pm
}^{\prime\prime}\right.  \left.  -\frac{4\pi^{2}}{k^{2}}\mathcal{L}_{\pm}%
^{2}\chi_{\pm}-\frac{1}{24}\chi_{\pm}^{\prime\prime\prime\prime}\right)
W_{\mp2}\nonumber\\
&  +\frac{1}{2}\left(  \chi_{\pm}^{\prime\prime}-\frac{8\pi}{k}\mathcal{L}%
_{\pm}\chi_{\pm}\right)  W_{0}\mp\frac{1}{6}\left(  \chi_{\pm}^{\prime
\prime\prime}-\frac{8\pi}{k}\chi_{\pm}\mathcal{L}_{\pm}^{\prime}-\frac{20\pi
}{k}\mathcal{L}_{\pm}\chi_{\pm}^{\prime}\right)  W_{\mp1}\ ,
\label{PTT-Lambda-W3}%
\end{align}
which depend on two arbitrary parameters per copy, $\varepsilon_{\pm}$,
$\chi_{\pm}$, being functions of $t$ and $\theta$, provided the transformation
law of the fields $\mathcal{L}_{\pm}$, $\mathcal{W}_{\pm}$ reads%
\begin{align}
\delta\mathcal{L}_{\pm}  &  =\varepsilon_{\pm}\mathcal{L}_{\pm}^{\prime
}+2\mathcal{L}_{\pm}\varepsilon_{\pm}^{\prime}-\frac{k}{4\pi}\varepsilon_{\pm
}^{\prime\prime\prime}-2\chi_{\pm}\mathcal{W}_{\pm}^{\prime}-3\mathcal{W}%
_{\pm}\chi_{\pm}^{\prime}\ ,\label{PTT-deltaL-W3}\\
\delta\mathcal{W}_{\pm}  &  =\varepsilon_{\pm}\mathcal{W}_{\pm}^{\prime
}+3\mathcal{W}_{\pm}\varepsilon_{\pm}^{\prime}-\frac{64\pi}{3k}\mathcal{L}%
_{\pm}^{2}\chi_{\pm}^{\prime}+3\chi_{\pm}^{\prime}\mathcal{L}_{\pm}%
^{\prime\prime}+5\mathcal{L}_{\pm}^{\prime}\chi_{\pm}^{\prime\prime}+\frac
{2}{3}\chi_{\pm}\mathcal{L}_{\pm}^{\prime\prime\prime}-\frac{k}{12\pi}%
\chi_{\pm}^{\prime\prime\prime\prime\prime}\nonumber\\
&  -\frac{64\pi}{3k}\left(  \chi_{\pm}\mathcal{L}_{\pm}^{\prime}-\frac
{5k}{32\pi}\chi_{\pm}^{\prime\prime\prime}\right)  \mathcal{L}_{\pm}\ .
\label{PTT-deltaW-W3}%
\end{align}
Then, the time component of the gauge fields $a_{t}^{\pm}$, is preserved under
the gauge transformations generated by (\ref{PTT-Lambda-W3}), with the
transformation rules in (\ref{PTT-deltaL-W3}), (\ref{PTT-deltaW-W3}), provided
the fields and the parameters are chiral:%
\begin{align}
\partial_{\mp}\mathcal{L}_{\pm}  &  =\partial_{\mp}\mathcal{W}_{\pm
}=0\ ,\label{PTT-FE-S-W3}\\
\partial_{\mp}\varepsilon_{\pm}  &  =\partial_{\mp}\chi_{\pm}=0\ .
\label{PTT-CC-S-W3}%
\end{align}
As in the case of pure gravity, the chirality of the fields in eq.
(\ref{PTT-FE-S-W3}) reflects the fact that the field equations in the
asymptotic region are satisfied.

The variation of the canonical generators that correspond to the asymptotic
symmetries spanned by (\ref{PTT-Lambda-W3}) now reads%
\begin{equation}
\delta Q_{\pm}\left(  \Lambda^{\pm}\right)  =-\frac{k}{2\pi}\int\left\langle
\Lambda^{\pm}\delta a_{\theta}^{\pm}\right\rangle d\theta=-\int\left(
\varepsilon_{\pm}\delta\mathcal{L}_{\pm}-\chi_{\pm}\delta\mathcal{W}_{\pm
}\right)  d\theta\ , \label{PTT-deltaQ-W3}%
\end{equation}
and then integrates as%
\begin{equation}
Q_{\pm}\left(  \Lambda^{\pm}\right)  =-\int\left(  \varepsilon_{\pm
}\mathcal{L}_{\pm}-\chi_{\pm}\mathcal{W}_{\pm}\right)  d\theta\ .
\label{PTT-Q-W3}%
\end{equation}
This means that generic gauge fields that fulfill the asymptotic conditions
described here, do not only carry spin-$2$ charges associated to
$\mathcal{L}_{\pm}$, whose zero modes are related to the energy and the
angular momentum, but they also possess spin-$3$ charges corresponding to
$\mathcal{W}_{\pm}$.

The algebra of the canonical generators can be straightforwardly recovered
from the transformation law of the fields in (\ref{PTT-deltaL-W3}),
(\ref{PTT-deltaW-W3}) and it is found to be given by two copies of the $W_{3}$
algebra with the same central extension as in pure gravity, i.e., $c=\frac
{3l}{2G}$. Once the fields are expanded in modes, the Poisson bracket algebra
is such that both copies fulfill
\begin{align}
i\left\{  \mathcal{L}_{m},\mathcal{L}_{n}\right\}   &  =\left(  m-n\right)
\mathcal{L}_{m+n}+\frac{k}{2}m^{3}\delta_{m+n,0}\ ,\nonumber\\
i\left\{  \mathcal{L}_{m},\mathcal{W}_{n}\right\}   &  =\left(  2m-n\right)
\mathcal{W}_{m+n}\ ,\label{PTT-W3-Algebra}\\
i\left\{  \mathcal{W}_{m},\mathcal{W}_{n}\right\}   &  =\frac{1}{3}\left(
m-n\right)  \left(  2m^{2}-mn+2n^{2}\right)  \mathcal{L}_{m+n}+\frac{16}%
{3k}\left(  m-n\right)  \Lambda_{m+n}+\frac{k}{6}m^{5}\delta_{m+n,0}%
\ ,\nonumber
\end{align}
where
\begin{equation}
\Lambda_{n}=\sum_{m}\mathcal{L}_{n-m}\mathcal{L}_{m}\ ,
\end{equation}
so that the algebra is manifestly nonlinear.

It has also been shown that once the asymptotic conditions
(\ref{PTT-amn-Standard W3}) are expressed in a suitable \textquotedblleft
decoupling\textquotedblright\ gauge choice, they admit a consistent vanishing
cosmological constant limit, so that the asymptotic symmetries are spanned by
a higher spin extension of the BMS$_{3}$ algebra with an appropriate central
extension \cite{PTT-GMPT} (see also \cite{PTT-Grumi-Flat}). Related results
along these lines, including Hamiltonian reduction \cite{PTT-HP-Flat},
unitarity \cite{PTT-GRR-Flat}, and the analysis of cosmologies endowed with
higher spin fields have been discussed in \cite{PTT-Chethan}, \cite{PTT-PZ},
\cite{PTT-Milne}, \cite{PTT-Grassmann-Flat}).

\subsection{Higher spin black hole proposal and its thermodynamics}

It is simple to verify that, for the case of constant functions $\mathcal{L}%
_{\pm}$ and $\mathcal{W}_{\pm}$, the asymptotic conditions described in the
previous subsection do not accommodate black holes carrying nontrivial
spin-$3$ charges. This is because once the holonomies along a thermal cycle
are required to be trivial, the spin-$3$ charges $\mathcal{W}_{\pm}$ are
forced to vanish. Thus, with the aim of finding black holes solutions which
could in principle be endowed with spin-$3$ charges, a different set of
asymptotic conditions was proposed in \cite{PTT-GK} (see section
\ref{PTT-Puzzles}) and further analyzed in \cite{PTT-CS}, \cite{PTT-CJS}.
Indeed, this set includes interesting new black holes solutions, which in the
static case are described by three constants, and the gauge fields are of the
form (\ref{PTT-Acontuti}), with%
\begin{align}
a^{\pm}  &  =\pm\left(  L_{\pm1}-\frac{2\pi}{k}\mathcal{\tilde{L}}L_{\mp1}%
\mp\frac{\pi}{2k}\mathcal{\tilde{W}}W_{\mp2}\right)  dx^{\pm}\nonumber\\
&  +\tilde{\mu}\left(  W_{\pm2}-\frac{4\pi}{k}\mathcal{\tilde{L}}W_{0}%
+\frac{4\pi^{2}}{k^{2}}\mathcal{\tilde{L}}^{2}W_{\mp2}\pm\frac{4\pi}%
{k}\mathcal{\tilde{W}}L_{\mp1}\right)  dx^{\mp}\ . \label{PTT-amn-GK}%
\end{align}
The precise form of the $SL\left(  3,%
\mathbb{R}
\right)  $ group elements $g_{\pm}=g_{\pm}\left(  \rho\right)  $, which was
further specified in \cite{PTT-AGKP}, would be needed in order to reconstruct
the metric and the spin-$3$ field according to eq. (\ref{PTT-metric+spin3}).
In the case of $sl\left(  3,%
\mathbb{R}
\right)  $ gauge fields, the conditions that guarantee the triviality of the
their holonomies around the thermal circle, since the representation in
(\ref{PTT-MR}) is vectorial, now read%
\begin{equation}
\mathcal{H}_{\tau}^{\pm}=e^{i\beta a_{t}^{\pm}}=\mathrm{1}\ ,
\label{PTT-Hat-GK}%
\end{equation}
which turn out to be equivalent to%
\begin{equation}
\mathrm{tr}\left[  \left(  a_{t}^{\pm}\right)  ^{3}\right]  =0\ \ ;\ \ \beta
^{2}\mathrm{tr}\left[  \left(  a_{t}^{\pm}\right)  ^{2}\right]  =8\pi^{2}\ .
\label{PTT-HOL-GK-Tr}%
\end{equation}
For the gauge fields (\ref{PTT-amn-GK}), conditions (\ref{PTT-HOL-GK-Tr})
reduce to%
\begin{align}
64\pi\mathcal{\tilde{L}}^{2}\tilde{\mu}\left(  32\pi\mathcal{\tilde{L}}%
\tilde{\mu}^{2}-9k\right)  +27k\mathcal{\tilde{W}}\left(  32\pi\mathcal{\tilde
{L}}\tilde{\mu}^{2}+k\right)  -864\pi k\mathcal{\tilde{W}}^{2}\tilde{\mu}^{3}
&  =0\ ,\label{PTT-Hat-1}\\
\frac{l^{2}\pi k}{2}\left(  \mathcal{\tilde{L}}-3\tilde{\mu}\mathcal{\tilde
{W}}+\frac{32\pi}{3k}\tilde{\mu}^{2}\mathcal{\tilde{L}}^{2}\right)  ^{-1}  &
=\beta^{2}\ , \label{PTT-Hat-2}%
\end{align}
respectively, which for the branch that is connected to the BTZ black hole,
being such that $\tilde{\mu}\rightarrow0$ when $\mathcal{\tilde{W}}%
\rightarrow0$, can be solved for $\beta$ and $\tilde{\mu}$ in terms of
$\mathcal{\tilde{L}}$ and $\mathcal{\tilde{W}}$, according to%
\begin{align}
\beta &  =\frac{l}{2}\sqrt{\frac{\pi k}{2\mathcal{\tilde{L}}}}\ \frac
{2C-3}{C-3}\left(  1-\frac{3}{4C}\right)  ^{-1/2}\ ,\label{PTT-betaGK}\\
\tilde{\mu}  &  =\frac{3}{4}\sqrt{\frac{kC}{2\pi\mathcal{\tilde{L}}}}%
\ \frac{1}{2C-3}\ , \label{PTT-muGK}%
\end{align}
where the constant $C$ is defined through%
\begin{equation}
\frac{C-1}{C^{3/2}}=\sqrt{\frac{k}{32\pi\mathcal{\tilde{L}}^{3}}%
}\mathcal{\tilde{W}\ }.
\end{equation}

A proposal to deal with the global charges and the thermodynamics of this
black hole solution, being based on a holographic approach, was put forward in
\cite{PTT-GK}, \cite{PTT-AGKP}. The bulk field equations were identified with
the Ward identities for the stress tensor and the spin-$3$ current of an
underlying dual CFT in two dimensions, so that the integration constant
$\mathcal{\tilde{L}}$ was interpreted as the stress tensor, while
$\mathcal{\tilde{W}}$ and $\tilde{\mu}$ were associated to the spin-$3$
current and its source, respectively. According to this prescription, the
first law of thermodynamics implies that the variation of the entropy should
be given by%
\begin{equation}
\delta\tilde{S}=\frac{4\pi}{l}\beta\left(  \delta\mathcal{\tilde{L}}%
-\tilde{\mu}\delta\mathcal{\tilde{W}}\right)  \ ,
\end{equation}
which by virtue of (\ref{PTT-betaGK}), (\ref{PTT-muGK}) integrates as%
\begin{equation}
\tilde{S}=4\pi\sqrt{2\pi k\mathcal{\tilde{L}}}\sqrt{1-\frac{3}{4C}}\ ,
\label{PTT-S-GK}%
\end{equation}
so that the trivial holonomy conditions around the thermal circle agree with
the integrability conditions of thermodynamics.

It is worth mentioning that the black hole entropy formula (\ref{PTT-S-GK})
remarkably agrees with the result found in \cite{PTT-GHJ}, which was obtained
from a completely different approach. Indeed, the computation of the free
energy was carried out directly in the dual CFT with extended conformal
symmetry in two dimensions, exploiting the properties of the partition
function under the S-modular transformation, making then no reference to the
holonomies in the bulk.

These approaches have been reviewed in \cite{PTT-Rew-GG}, \cite{PTT-Rew-AGKP},
\cite{PTT-Rew-J}, and further results about black hole thermodynamics along
these lines have been found in \cite{PTT-F-GK1}, \cite{PTT-F-GK2},
\cite{PTT-F-GK3z}, \cite{PTT-KU}, \cite{PTT-dBJ}, \cite{PTT-dBJ2},
\cite{PTT-F-GK3}, \cite{PTT-F-GK4}, \cite{PTT-Last-1}, \cite{PTT-Last}.

\bigskip

However, it should be stressed that identifying the integration constants
$\mathcal{\tilde{L}}$ and $\mathcal{\tilde{W}}$ with global charges, appears
to be very counterintuitive from the point of view of the canonical formalism.
This is because, in spite of the fact that the components of the gauge fields
along $dx^{\pm}$ for the black hole solution (\ref{PTT-amn-GK}) agree with the
ones of the asymptotic fall-off in (\ref{PTT-amn-Standard W3}), once a
nonvanishing constant $\tilde{\mu}$ is included, the additional terms along
$dx^{\mp}$ amount to a severe modification of the asymptotic form of the
dynamical fields $a_{\theta}^{\pm}$, so that the expression for the global
charges in eq. (\ref{PTT-Q-W3}) no longer applies for this class of black hole
solutions. Hence, as shown in \cite{PTT-PTT1}, in full analogy with what
occurs in the case of three-dimensional General Relativity coupled to scalar
fields with slow fall-off at infinity \cite{PTT-HMTZ-2+1}, \cite{PTT-HMTZ-Log}%
, the effect of modifying the asymptotic behaviour is such that the total
energy acquires additional nonlinear contributions in the deviation of the
fields with respect to the reference background. Indeed, the variation of the
total energy can be obtained directly from (\ref{PTT-deltaE}), which for the
case of the black hole solution (\ref{PTT-amn-GK}), reads%
\begin{align}
\delta E  &  =\frac{k}{2\pi}\int\left(  \left\langle a_{t}^{+}\delta
a_{\theta}^{+}\right\rangle -\left\langle a_{t}^{-}\delta a_{\theta}%
^{-}\right\rangle \right)  d\theta\ ,\nonumber\\
&  =\frac{4\pi}{l}\left[  \delta\mathcal{\tilde{L}}-\frac{32\pi}{3k}%
\delta(\mathcal{\tilde{L}}^{2}\mu^{2})+\tilde{\mu}\delta\mathcal{\tilde{W}%
}+3\mathcal{\tilde{W}}\delta\tilde{\mu}\right]  \ . \label{PTT-deltaE1-GK}%
\end{align}

Note that (\ref{PTT-deltaE1-GK}) is not an exact differential. This is natural
because the variation of the total energy not only includes the variation of
the mass, but also the contribution coming from all the constraints.
Therefore, in order to suitably disentangle the mass (internal energy) from
the work terms, one should provide a consistent set of asymptotic conditions
that allows the precise identification of the global charges as well as the
chemical potentials. This is discussed in section \ref{PTT-Puzzles}.
Nonetheless, the expression (\ref{PTT-deltaE1-GK}) provides a nice shortcut to
compute the black hole entropy, circumventing the explicit computation of
higher spin charges and their chemical potentials \cite{PTT-PTT1},
\cite{PTT-PTT2}. This is because, by virtue of the first law, the inverse
temperature $\beta$ acts as an integrating factor, being such that the product
$\beta\delta E$ becomes an exact differential that corresponds to the
variation of the entropy, i.e.,%
\begin{equation}
\delta S=\beta\delta E=\delta\left[  4\pi\sqrt{2\pi k\mathcal{\tilde{L}}%
}\left(  1-\frac{3}{2C}\right)  ^{-1}\sqrt{1-\frac{3}{4C}}\right]  \ ,
\end{equation}
so that the black hole entropy is given by%
\begin{equation}
S=4\pi\sqrt{2\pi k\mathcal{\tilde{L}}}\left(  1-\frac{3}{2C}\right)
^{-1}\sqrt{1-\frac{3}{4C}}\ . \label{PTT-Sc-GK}%
\end{equation}

\bigskip

As explained in \cite{PTT-PTT2}, the entropy (\ref{PTT-Sc-GK}) can be
recovered from a suitable generalization of the Bekenstein-Hawking formula,
given by
\begin{equation}
S=\frac{A}{4G}\cos\left[  \frac{1}{3}\arcsin\left(  3^{3/2}\frac{\varphi_{+}%
}{A^{3}}\right)  \right]  \ , \label{PTT-EntropyHS}%
\end{equation}
which depends on the reparametrization invariant integrals of the pullback of
the metric and the spin-$3$ field at the spacelike section of the horizon,
i.e., on the horizon area $A$ and its spin-$3$ analogue:%
\begin{equation}
\varphi_{+}^{1/3}:=\int_{\partial\Sigma_{+}}\left(  \varphi_{\mu\nu\rho}%
\frac{dx^{\mu}}{d\sigma}\frac{dx^{\nu}}{d\sigma}\frac{dx^{\rho}}{d\sigma
}\right)  ^{1/3}d\sigma\ .
\end{equation}

It is worth highlighting that, for the static case, and in the weak spin-$3$
field limit, our expression for the entropy (\ref{PTT-EntropyHS}) reduces to%
\begin{equation}
S=\frac{A}{4G}\left.  \left(  1-\frac{3}{2}\left(  g^{\theta\theta}\right)
^{3}\varphi_{\theta\theta\theta}^{2}+\mathcal{O}\left(  \varphi^{4}\right)
\right)  \right\vert _{\rho_{+}}\ , \label{PTT-Entropy-pert}%
\end{equation}
in full agreement with the result found in \cite{PTT-CFPT2}, which was
obtained from a completely different approach. Indeed, in \cite{PTT-CFPT2} the
action was written in terms of the metric and the perturbative expansion of
the spin-$3$ field up to quadratic order, so that the correction to the area
law in (\ref{PTT-Entropy-pert}) was found by means of Wald's formula
\cite{PTT-Wald-Entropy}.

Further results about black hole thermodynamics and along these lines have
been found in \cite{PTT-dBJ}, \cite{PTT-dBJ2}, \cite{PTT-ACI}, \cite{PTT-V},
\cite{PTT-LLW}, \cite{PTT-F-PTT1}, and the variation of the total energy
(\ref{PTT-deltaE1-GK}) has also been recovered through different methods in
\cite{PTT-CS}, \cite{PTT-CJS}.

\bigskip

Since the entropy is expected to be an intrinsic property of the black hole,
the fact that the nonperturbative expression for the entropy $S$ in eq.
(\ref{PTT-Sc-GK}) differs from $\tilde{S}$ in (\ref{PTT-S-GK}) by a factor
that characterizes the presence of the spin-three field, i.e.,$\ S=\tilde
{S}\left(  1-\frac{3}{2C}\right)  ^{-1}$, is certainly disturbing. Indeed,
curiously, a variety of different approaches either lead to $\tilde{S}$ or
$S$, in refs. \cite{PTT-GK}, \cite{PTT-GHJ}, \cite{PTT-KU}, \cite{PTT-Last-1},
\cite{PTT-Last}, and \cite{PTT-PTT2}, \cite{PTT-ACI}, \cite{PTT-V},
\cite{PTT-LLW}, respectively, or even to both results \cite{PTT-dBJ},
\cite{PTT-dBJ2} for the black hole entropy.

\bigskip

As explained in \cite{PTT-PTT1}, \cite{PTT-PTT2}, the discrepancy of these
results stems from the mismatch in the definition of global charges
aforementioned, which turns out to be inherited by the entropy once computed
through the first law, even in the weak spin-$3$ field limit.

\bigskip

Nonetheless, some puzzles still remain to be clarified, as it is the question
about how the entropy (\ref{PTT-Sc-GK}) fulfills the first law of
thermodynamics in the grand canonical ensemble, which is related to whether
the black hole solution (\ref{PTT-amn-GK}) actually carries or not a global a
spin-$3$ charge. This is discussed in the next section \ref{PTT-Puzzles}.

\section{Solving the puzzles: asymptotic conditions revisited and different
classes of black holes}

\label{PTT-Puzzles}

As explained in \cite{PTT-HPTT}, \cite{PTT-BHPTT}, the puzzles mentioned above
become resolved once the asymptotic conditions are extended so as to admit a
generic choice of chemical potentials associated to the higher spin charges,
so that the original asymptotic $W_{3}$ symmetries are manifestly preserved by
construction. In this way, any possible ambiguity is removed. This can be seen
as follows. At a slice of fixed time, according to (\ref{PTT-amn-Standard W3}%
), the asymptotic behaviour of the dynamical fields is of the form%
\begin{equation}
a_{\theta}^{\pm}=\left(  L_{\pm1}-\frac{2\pi}{k}\mathcal{L}_{\pm}L_{\mp
1}-\frac{\pi}{2k}\mathcal{W}_{\pm}W_{\mp2}\right)  d\theta\ ,
\label{PTT-a-theta-W3}%
\end{equation}
which is maintained under the gauge transformations $\Lambda^{\pm}$, defined
through (\ref{PTT-Lambda-W3}), with (\ref{PTT-deltaL-W3})
and\ (\ref{PTT-deltaW-W3}). In order to determine the asymptotic form of the
gauge fields along time evolution, note that the field equations $F_{ti}=0$
read%
\[
\dot{A}_{i}=\partial_{i}A_{t}+\left[  A_{i},A_{t}\right]  \ ,
\]
which implies that the time evolution of the dynamical fields corresponds to a
gauge transformation parametrized by $A_{t}$. Hence, in order to preserve the
asymptotic symmetries along the evolution in time, the Lagrange multipliers
must be of the allowed form (\ref{PTT-Lambda-W3}), i.e., $a_{t}^{\pm}%
=\Lambda^{\pm}$. Thus, following \cite{PTT-HPTT}, the chemical potentials are
included in the time component of the gauge fields only, so that the
asymptotic form of the gauge fields is given by
\begin{equation}
a^{\pm}=\pm\left(  L_{\pm1}-\frac{2\pi}{k}\mathcal{L}_{\pm}L_{\mp1}-\frac{\pi
}{2k}\mathcal{W}_{\pm}W_{\mp2}\right)  dx^{\pm}\pm\frac{1}{l}\Lambda^{\pm}%
(\nu_{\pm},\mu_{\pm})dt\ , \label{PTT-amn-pot-w3}%
\end{equation}
where $\nu_{\pm}$, $\mu_{\pm}$ stand for arbitrary fixed functions of $t$,
$\theta$ without variation ($\delta\nu_{\pm}=\delta\mu_{\pm}=0$), that
correspond to the chemical potentials. Note that, since the asymptotic form of
the dynamical fields (\ref{PTT-a-theta-W3}) is unchanged as compared with
(\ref{PTT-amn-Standard W3}), the expression for the global charges remains the
same, i.e., at a fixed $t$ slice, the global charges are again given by
(\ref{PTT-Q-W3}), so that the asymptotic symmetries are still generated by two
copies of the $W_{3}$ algebra.

Consistency then requires that the asymptotic form of $a_{t}^{\pm}$, should
also be preserved under the asymptotic symmetries, which implies that the
field equations have to be fulfilled in the asymptotic region, and the
parameters of the asymptotic symmetries satisfy \textquotedblleft deformed
chirality conditions\textquotedblright, which read%
\begin{align}
l\mathcal{\dot{L}}_{\pm}  &  =\pm\left(  1+\nu_{\pm}\right)  \mathcal{L}_{\pm
}^{\prime}\mp2\mu_{\pm}\mathcal{W}_{\pm}^{\prime}\ ,\nonumber\\
l\mathcal{\dot{W}}_{\pm}  &  =\pm\left(  1+\nu_{\pm}\right)  \mathcal{W}_{\pm
}^{\prime}\pm\frac{2}{3}\mu_{\pm}\left(  \mathcal{L}_{\pm}^{\prime\prime
\prime}-\frac{16\pi}{k}\left(  \mathcal{L}_{\pm}^{2}\right)  ^{\prime}\right)
\ , \label{PTT-WLpunto}%
\end{align}
and%
\begin{align}
l\dot{\chi}_{\pm}  &  =\pm\left(  1+\nu_{\pm}\right)  \chi_{\pm}^{\prime}%
\pm2\mu_{\pm}\varepsilon_{\pm}^{\prime}\ ,\nonumber\\
l\dot{\varepsilon}_{\pm}  &  =\pm\left(  1+\nu_{\pm}\right)  \varepsilon_{\pm
}^{\prime}\mp\frac{2}{3}\mu_{\pm}\left(  \chi_{\pm}^{\prime\prime\prime}%
-\frac{32\pi}{k}\chi_{\pm}^{\prime}\mathcal{L}_{\pm}\right)  \ ,
\label{PTT-EpsilonXipunto}%
\end{align}
respectively, where for simplicity, in eqs. (\ref{PTT-WLpunto}),
(\ref{PTT-EpsilonXipunto}), the chemical potentials associated to the spin-$2$
and spin-$3$ charges, given by $\nu_{\pm}$ and $\mu_{\pm}$, were assumed to be constants.

\bigskip

Therefore, by construction, the functions $\mathcal{L}_{\pm}$, $\mathcal{W}%
_{\pm}$ are really what they mean, since their Poisson brackets fulfill the
$W_{3}$ algebra with the same central extension. Note that this is so
regardless the choice of chemical potentials, because the canonical generators
do no depend on the Lagrange multipliers.

The asymptotic conditions given by (\ref{PTT-amn-pot-w3}) then provide the
required extension of the ones in \cite{PTT-Henneaux-Rey}, \cite{PTT-CFPT1},
since the latter are recovered when the chemical potentials are switched off,
i.e., for $\nu_{\pm}=0$, $\mu_{\pm}=0$. In this case, eqs. (\ref{PTT-WLpunto})
and (\ref{PTT-EpsilonXipunto}) reduce to (\ref{PTT-FE-S-W3}) and
(\ref{PTT-CC-S-W3}), respectively, expressing the fact that the fields and the
parameters become chiral.

From a different perspective, the case of $\nu_{\pm}=-1$, $\mu_{\pm}=1$ has
also been discussed in \cite{PTT-G-Lif}.

\bigskip

It is worth emphasizing that since the Lagrange multipliers appear in the
improved action through the improved generators (\ref{PTT-improved-G}), the
interpretation of $\nu_{\pm}$, $\mu_{\pm}$ as chemical potentials, is also
guaranteed by construction. Note that this corresponds to the standard
procedure one follows in the case of Reissner-Nordström black holes, where the
chemical potential associated to the electric charge corresponds to the time
component of the electromagnetic field, being the Lagrange multiplier of the
$U(1)$ constraint.

\bigskip

The extended asymptotic conditions (\ref{PTT-amn-pot-w3}), in the case of
constant functions $\mathcal{L}_{\pm}$, $\mathcal{W}_{\pm}$ and chemical
potentials $\nu_{\pm}$, $\mu_{\pm}$, then accommodate a new class of black
hole solutions, endowed not only with with mass and angular momentum, but also
with nontrivial well-defined spin-$3$ charges \cite{PTT-HPTT}. Their
asymptotic and thermodynamical properties are further discussed in
\cite{PTT-BHPTT}, where it is explicitly shown that for this solution, there
is no tension between the different approaches mentioned above.

Note that in the standard approach for black hole thermodynamics, the
temperature and the chemical potential for the angular momentum do not
explicitly appear in the fields. Instead, they enter through the
identifications involving the Euclidean time and the angle, so that the range
of the coordinates is not fixed and depends on the solution. The presence of
nonvanishing chemical potentials $\nu_{\pm}$ associated to the spin-$2$
charges, then allows performing the description keeping the range of the
coordinates fixed once and for all, i.e., $0\leq\theta<2\pi$ and $0\leq
\tau<2\pi l$, which amounts to introduce a non trivial lapse and shift in the
metric formalism. Both approaches are indeed equivalent, but in the case of
higher spin black holes, since the chemical potentials that correspond to the
spin-$3$ charges cannot be absorbed into the modular parameter of the torus,
it becomes conceptually safer to follow the latter approach, since all the
chemical potentials become introduced and treated unambiguously in the same footing.

Otherwise, for instance, if the chemical potentials were not introduced along
the thermal circles, but instead along additional non-vanishing components of
the gauge fields along the conjugate null directions, as in the case of
\cite{PTT-GK}, the asymptotic form of the gauge fields would be given by%
\begin{equation}
a^{\pm}=\pm\left(  L_{\pm1}-\frac{2\pi}{k}\mathcal{\tilde{L}}_{\pm}L_{\mp
1}-\frac{\pi}{2k}\mathcal{\tilde{W}}_{\pm}W_{\mp2}\right)  dx^{\pm}\pm
\Lambda^{\pm}\left(  \tilde{\nu}_{\pm},\tilde{\mu}_{\pm}\right)  dx^{\mp},
\end{equation}
which severely modifies the components of the dynamical fields $a_{\theta
}^{\pm}$, in a way that is incompatible with the asymptotic $W_{3}$ symmetry.
This is because at a fixed $t$ slice, the terms proportional to $\tilde{\mu
}_{\pm}$ contribute to $a_{\theta}^{\pm}$ with additional terms of the form%
\begin{align}
a_{\theta}^{\pm}  &  =\left(  L_{\pm1}-\frac{2\pi}{k}\mathcal{\tilde{L}}_{\pm
}L_{\mp1}-\frac{\pi}{2k}\mathcal{\tilde{W}}_{\pm}W_{\mp2}\right)  +(\tilde
{\nu}_{\pm}L_{\pm1}+\tilde{\mu}_{\pm}W_{\pm2})\nonumber\\
&  +\left[  \frac{1}{2}\left(  -\frac{4\pi}{k}\tilde{\nu}_{\pm}\mathcal{\tilde
{L}}_{\pm}+\frac{8\pi}{k}\mathcal{\tilde{W}}_{\pm}\tilde{\mu}_{\pm}\right)
L_{\mp1}-\left(  \frac{\pi}{2k}\mathcal{\tilde{W}}_{\pm}\tilde{\nu}_{\pm
}-\frac{4\pi^{2}}{k^{2}}\mathcal{\tilde{L}}_{\pm}^{2}\tilde{\mu}_{\pm}\right)
W_{\mp2}\right] \nonumber\\
&  -\frac{4\pi}{k}\mathcal{\tilde{L}}_{\pm}\tilde{\mu}_{\pm}W_{0}\ ,
\end{align}
that are not of highest (or lowest) weight, and hence incompatible with the
asymptotic conditions (\ref{PTT-amn-pot-w3}) that implement the Hamiltonian
reduction of the current algebra associated to $sl(3,%
\mathbb{R}
)$ to the $W_{3}$ algebra. Indeed, in this case, the asymptotic symmetries
that preserve the asymptotic form of $a_{\theta}$ are shown to be spanned by
two copies of the Bershardsky-Polyakov algebra $W_{3}^{2}$ \cite{PTT-Polyakov}%
, \cite{PTT-Bershadsky}, corresponding to the other non trivial (so-called
diagonal) embedding of $sl\left(  2,%
\mathbb{R}
\right)  $ into $sl\left(  3,%
\mathbb{R}
\right)  $ \cite{PTT-BHPTT}. Therefore, in spite of dealing with the same
action, the effect of this drastic modification of the boundary conditions
amounts to deal with a completely different theory, being characterized by a
different field content, and hence with an inequivalent spectrum, so that
their corresponding black hole solutions, as the one in (\ref{PTT-amn-GK}),
are characterized by another set of global charges of lower spin.

\bigskip

It is worth pointing out that our procedure to incorporate chemical potentials
can be straightforwardly extended to the case of $\mathfrak{g}_{\pm}=sl\left(
N,%
\mathbb{R}
\right)  $, regardless the way in which $sl\left(  2,%
\mathbb{R}
\right)  $ is embedded, as well as to the case of infinite-dimensional higher
spin algebras.

\bigskip

Some closing remarks are in order. It should be mentioned that the case of
three-dimensional gravity nonminimally coupled with spin-$3$ fields, also
appears to be consistently formulated in the second-order formalism by
introducing a suitable set of auxiliary fields \cite{PTT-FN}. Besides, in the
case of spin-$3$ and higher, consistent sets of asymptotic conditions have
also been proposed in \cite{PTT-Henneaux-Rey}, \cite{PTT-CFP}, \cite{PTT-GH},
while exact solutions and their properties have been explored in
\cite{PTT-PK}, \cite{PTT-TAN}, \cite{PTT-GGR}, \cite{PTT-BCT}, \cite{PTT-FHK}.
In the context of higher spin supergravity in three dimensions, the asymptotic
structure was analyzed in \cite{PTT-Super-HS}, and exact solutions have also
been found in \cite{PTT-Super-1}, \cite{PTT-Super-SD}, \cite{PTT-Super-T}.
Moreover, along the lines of holography and the corresponding dual CFT theory
with extended conformal symmetry at the boundary \cite{PTT-GGS},
\cite{PTT-H1}, \cite{PTT-H2}, further interesting results can also be found in
\cite{PTT-CGGR}, \cite{PTT-H4}, \cite{PTT-H5}, \cite{PTT-H6}, \cite{PTT-CPR},
\cite{PTT-H7}, \cite{PTT-GJP}, \cite{PTT-CF}.

\acknowledgments We thank G. Barnich, X. Bekaert, E. Bergshoeff, C. Bunster,
A. Campoleoni, R. Canto, D. Grumiller, M. Henneaux, J. Jottar, C. Martínez, J.
Matulich, J. Ovalle, R. Rahman, S-J Rey, J. Rosseel, C. Troessaert and M.
Vasiliev for stimulating discussions. R.T. also thanks L. Papantonopoulos and
the organizers of the Seventh Aegean Summer School, \textquotedblleft Beyond
Einstein's theory of gravity\textquotedblright, for the opportunity to give
this lecture in a wonderful atmosphere. This work is partially funded by the
Fondecyt grants N${^{\circ}}$ 1130658, 1121031, 11130260, 11130262. The Centro
de Estudios Científicos (CECs) is funded by the Chilean Government through the
Centers of Excellence Base Financing Program of Conicyt.

\end{document}